\documentclass{PoS}

\title{Transverse-momentum resummation for gaugino-pair production
at the LHC}

\ShortTitle{Transverse-momentum resummation for gaugino-pair production
at the LHC}

\author{\speaker{Jonathan DEBOVE}%
  \\
  Laboratoire de Physique Subatomique et de Cosmologie,
  Universit\'e Joseph Fourier/CNRS-IN2P3/INPG,
  53 Avenue des Martyrs, F-38026 Grenoble, France \\
  E-mail: \email{debove@lpsc.in2p3.fr}}

\abstract{
  We present a first precision analysis of the transverse-momentum 
  spectrum of gaugino pairs produced at the LHC with center-of-mass energies 
  of 10 or 14 TeV.
  Our calculation is based on a universal resummation formalism at 
  next-to-leading logarithmic accuracy, which is consistently matched 
  to the perturbative prediction at $O(\alpha_s)$. 
  Numerical results are given for the ``gold-plated'' associated production 
  of neutralinos and charginos for a typical benchmark point 
  in the constrained Minimal Supersymmetric Standard Model. 
  We show that the matched resummation results 
  differ considerably from the Monte Carlo predictions employed traditionally 
  in experimental analyses.
  We also investigate in detail the theoretical uncertainties coming from
  scale and parton-density function variations and non-perturbative effects.}

\FullConference{European Physical Society Europhysics Conference 
  on High Energy Physics\\ July 16-22, 2009\\ Krakow, Poland}

\begin{document}  

  \section{Introduction}
  The Minimal Supersymmetric Standard Model (MSSM) is one of the most 
  appealing extensions of the Standard Model of particle physics 
  \cite{Nilles:1983ge,Haber:1984rc}.
  In the MSSM, the fermionic partners of the neutral (charged) gauge 
  and Higgs bosons are called neutralinos (charginos) 
  and are of particular importance. 
  Stabilized by $R$-symmetry, the lightest neutralino 
  is a very promising candidate for the dark matter 
  observed in the universe.
  Moreover, the values of gaugino masses and mixings 
  are key ingredients to understand both the electroweak  
  symmetry and the supersymmetry breakings. 
  Since these particles may be light enough to be produced at current 
  hadron colliders, they have been the object of particular attention, 
  and cross sections for the production of gaugino pairs have been 
  extensively studied at leading order 
  \cite{Barger:1983wc, Dawson:1983fw, Bozzi:2007me, Debove:2008nr, Fuks:2008ab}
  and next-to-leading order of perturbative QCD \cite{Beenakker:1999xh}.
 
  While particle pairs are produced with zero transverse momentum $(p_T)$
  in the Born approximation, gluon bremsstrahlung induces non-zero $p_T$
  at $O(\alpha_s)$ in the strong coupling constant.
  The aim of this work is to perform an accurate calculation of the 
  $p_T$-distribution of the gaugino pairs.
  Although the use of the perturbative expansion in powers of $\alpha_s$ 
  is fully justified, when the $p_T$ of the produced system is of the order 
  of its invariant mass $M$, the convergence of the perturbative 
  expansion is spoiled by powers of large logarithmic terms, $\ln M^2/p_T^2$,
  in the region where $p_T \ll M$.
  Therefore the enhanced logarithms must be resummed to all orders in 
  $\alpha_s$.

  \section{Transverse-momentum resummation}
  The method to systematically perform all-order resummation of classes of 
  enhanced logarithms is well-known 
  \cite{Collins:1984kg,
    Bozzi:2005wk} 
  and is performed in impact parameter ($b$) space.
  Thus we work with the Fourier transform $W$ of the partonic cross section
  defined by 
  \begin{equation}
    \frac{d \sigma^{RES}}{d M^2 d p_T^2} (z) = \frac{z}{M^2} 
    \int \frac{d^2 b}{4 \pi} \; e^{i \mathbf{b} \cdot \mathbf{p_T}} \; 
    W(b^2, M^2, z)\,,
  \end{equation}
  where $z=M^2/s$ and $\sqrt{s}$ is the partonic center-of-mass energy.
  The renormalisation scale ($\mu_R$) and the factorisation scale ($\mu_F$)
  dependencies have been removed for the sake of simplicity.
  After a Mellin transform with resptect to $z$, the $N$-moments
  of the $W$-function
  \begin{equation}
    W(b^2, M^2, N)=H(M^2, N)\exp[G(b^2, N)]
  \end{equation}
  factorize into a $b$-independent function $H$ and 
  an exponential form factor $\exp [G]$.
  The function $H$ encodes the full process dependence, 
  and the form factor $\exp[G]$ resums all the terms that are 
  order by order logarithmically divergent.
  The exponent $G$, which controls the soft gluon emmission, is thus 
  universal, i.e. it does not depend on the final state of the process
  \cite{
    Bozzi:2005wk}.
  The general expressions for the $H$- and $G$-functions can be found 
  in Refs. \cite{Bozzi:2005wk, Debove:2009ia}.


  After the large logarithms have been resummed in $b$-space, 
  we have to switch back to $p_T$-space in order to achieve 
  a phenomenological study.
  Special attention has to be paid to the singularities 
  in the exponent $G$.
  They are related to the presence of the Landau pole in the 
  perturbative running of $\alpha_s$, and a prescription is needed.
  In our numerical study, we follow Ref. \cite{Laenen:2000de} and deform 
  the integration contour in the complex $b$-plane.
  Note that the divergent behavior of $\alpha_s$ signals the onset of 
  non-perturbative (NP) effects at large $b$. These unknown effects may 
  be extracted from experiment and can then be included in the 
  exponential form factor.
  
  Finally, in order to conserve the full information contained in the 
  fixed-order calculation, the $O(\alpha_s)$ and the resummed 
  calculations are matched by subtracting from their sum 
  the resummed cross section truncated at $O(\alpha_s)$.
  
  \section{Numerical results}

  We now present numerical results for the associated production 
  of neutralinos and charginos at the LHC with a hadronic 
  center-of-mass energy of $\sqrt{S}=$10 and 14 TeV.
  Results for the pair production of neutralinos and charginos
  at the Tevatron and at the LHC can be found in Ref. \cite{Debove:2009ia}.
  The parton densities are evaluated in the most recent parametrisation
  of the CTEQ collaboration \texttt{CTEQ6.6M} \cite{Nadolsky:2008zw}
  with $\mu_F$ (and $\mu_R$) set to the average mass $m_{\tilde{\chi}}$
  of the final state particles, 
  and $\alpha_s$ is evaluated at two-loop accuracy.
  In the following, we choose the minimal supergravity benchmark point 
  SPS1a' \cite{AguilarSaavedra:2005pw} 
  and obtain the weak-scale supersymmetric parameters through 
  the computer code 
  \texttt{SuSpect2.3} \cite{Djouadi:2002ze}.

  \begin{figure}
    \centering
    \begin{minipage}[t]{.49\columnwidth}
      \includegraphics[width=\columnwidth]{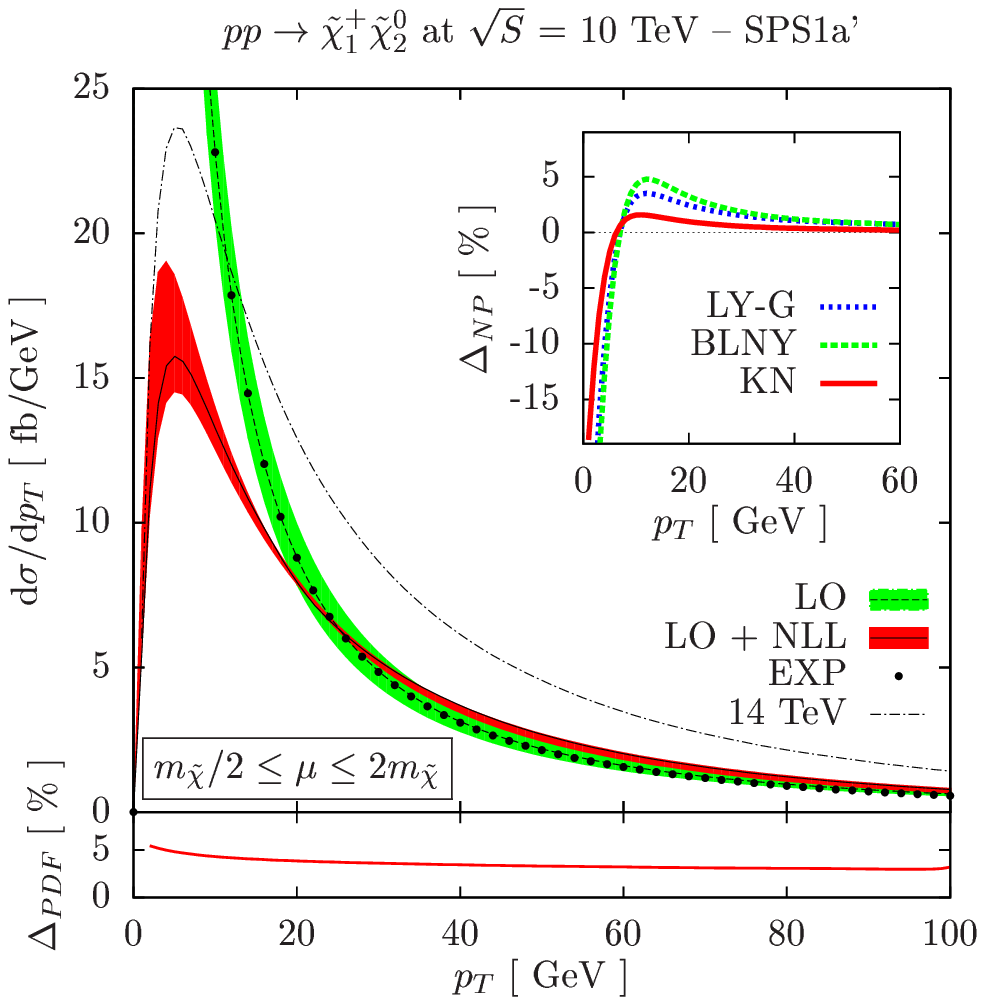}
      \caption{	$p_T$-spectra of 
	$\tilde{\chi}_1^+ \tilde{\chi}_2^0$-pairs at the LHC.
	The LO calculation (dashed) is matched to the resummed
	calculation (full) by subtracting its fixed-order expansion
	(dotted).
	The scale (shaded band), PDF (below) and NP (insert) 
	uncertainties are shown, as well as the matched result 
	for the LHC design energy of 14 TeV (dot-dashed).}
      \label{fig:1}
    \end{minipage}\hfill
    \begin{minipage}[t]{.49\columnwidth}
      \includegraphics[width=\columnwidth]{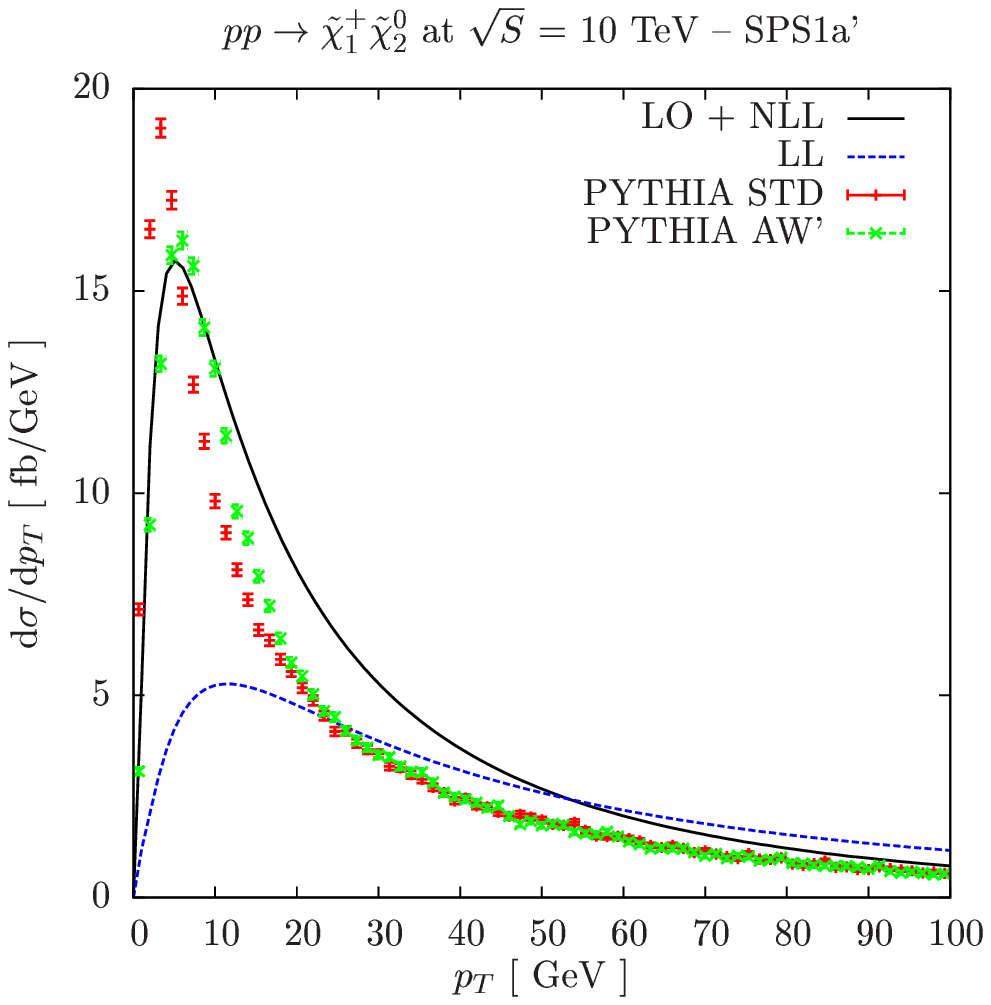}
      \caption{$p_T$-spectra of 
	$\tilde{\chi}_1^+ \tilde{\chi}_2^0$-pairs at the LHC.
	The matched LO+NLL (full) and the LL (dashed) results
	are compared with the predictions of the PYTHIA 
	parton shower with default (bars) and tuned (crosses) 
	parameters.}
      \label{fig:2}
    \end{minipage}
  \end{figure}
  In Fig. \ref{fig:1}, we show the $p_T$-spectra of chargino-neutralino 
  pairs produced at the Tevatron and at the LHC. 
  As expected, the $O(\alpha_s)$ calculation (LO) diverges at low $p_T$, 
  but becomes finite after having been matched to the resummed prediction 
  at next-to-leading logarithmic (LO+NLL) accuracy. 
  In this region the perturbative expansion (EXP) of the resummed 
  prediction coincides with the LO one.
  For comparison, the LO+NLL prediction for the 14 TeV design energy 
  of the LHC is also presented.
  We also study three different sources of uncertainty:
  the scale variations evaluated in the range 
  $[m_{\tilde{\chi}}/2, \,  2m_{\tilde{\chi}}]$,
  the PDF uncertainties $\Delta_{PDF}$ as defined by the CTEQ collaboration
  \cite{Nadolsky:2008zw} and the choice of three parametrisations
  for the NP form factor evaluated through
  $\Delta_{NP}=(d\sigma^{NP}-d\sigma)/d\sigma$
  \cite{Landry:2002ix, Konychev:2005iy}.
  The scale dependence of the LO+NLL prediction is clearly improved 
  with respect to the LO result and the other uncertainties are all smaller 
  than 5\% for $p_T>5$ GeV.
  
  In Fig. \ref{fig:2}, we compare our LO+NLL prediction with our 
  resummed result at leading-logarith\-mic (LL) accuracy and two 
  different setups for the \texttt{PYTHIA6.4} \cite{Sjostrand:2006za}
  Monte Carlo (MC) generator.
  We see that the default (STD) MC simulation is clearly improved beyond the 
  LL approximation and approaches the LO+NLL result, but peaks at slightly 
  smaller values of $p_T$.
  This behavior can be improved by tuning the intrinsic $p_T$ of the partons
  in the hadron (AW') \cite{Debove:2009ia},  
  but both simulations underestimate the intermediate $p_T$-region.


  \section{Conclusion}
  In summary, we have calculated the $p_T$-distribution of 
  gaugino pairs produced at the LHC at next-to-leading 
  logarithmic accuracy. 
  This has the advantage of making the perturbative predictions finite
  and reducing the scale uncertainties.
  We have also compared our results with a MC generator commonly used 
  for experimental analyses.
  All these features will possibly lead to improvements for the 
  experimental determination of the gaugino parameters.

  

\begin{thebibliography}{99}    


  \bibitem{Nilles:1983ge}
    H.~P.~Nilles,
    Phys.\ Rept.\  {\bf 110} (1984) 1.

  \bibitem{Haber:1984rc}
    H.~E.~Haber and G.~L.~Kane,
    Phys.\ Rept.\  {\bf 117} (1985) 75.


  \bibitem{Barger:1983wc}
    V.~D.~Barger, R.~W.~Robinett, W.~Y.~Keung and R.~J.~N.~Phillips,
    Phys.\ Lett.\  B {\bf 131} (1983) 372.

  \bibitem{Dawson:1983fw}
    S.~Dawson, E.~Eichten and C.~Quigg,
    Phys.\ Rev.\  D {\bf 31} (1985) 1581.

  \bibitem{Bozzi:2007me}
    G.~Bozzi, B.~Fuks, B.~Herrmann and M.~Klasen,
    Nucl.\ Phys.\  B {\bf 787} (2007) 1.

  \bibitem{Debove:2008nr}
    J.~Debove, B.~Fuks and M.~Klasen,
    Phys.\ Rev.\  D {\bf 78} (2008) 074020.

  \bibitem{Fuks:2008ab}
    B.~Fuks, B.~Herrmann and M.~Klasen,
    Nucl.\ Phys.\  B {\bf 810} (2009) 266.
    
  \bibitem{Beenakker:1999xh}
    W.~Beenakker, M.~Klasen, M.~Kr\"amer, T.~Plehn, M.~Spira and P.~M.~Zerwas,
    Phys.\ Rev.\ Lett.\  {\bf 83} (1999) 3780
    [Erratum-ibid.\  {\bf 100} (2008) 029901].



  \bibitem{Collins:1984kg}
    J.~C.~Collins, D.~E.~Soper and G.~Sterman,
    Nucl.\ Phys.\  B {\bf 250} (1985) 199.


  \bibitem{Bozzi:2005wk}
    G.~Bozzi, S.~Catani, D.~de Florian and M.~Grazzini,
    Nucl.\ Phys.\  B {\bf 737} (2006) 73.
    
  \bibitem{Debove:2009ia}
    J.~Debove, B.~Fuks and M.~Klasen,
    arXiv:0907.1105 [hep-ph].

  \bibitem{Laenen:2000de}
    E.~Laenen, G.~Sterman and W.~Vogelsang,
    Phys.\ Rev.\ Lett.\  {\bf 84} (2000) 4296.

  \bibitem{Nadolsky:2008zw}
    P.~M.~Nadolsky {\it et al.},
    Phys.\ Rev.\  D {\bf 78} (2008) 013004.

  \bibitem{AguilarSaavedra:2005pw}
    J.~A.~Aguilar-Saavedra {\it et al.},
    Eur.\ Phys.\ J.\  C {\bf 46} (2006) 43.

  \bibitem{Djouadi:2002ze}
    A.~Djouadi, J.~L.~Kneur and G.~Moultaka,
    Comput.\ Phys.\ Commun.\  {\bf 176} (2007) 426.


  \bibitem{Landry:2002ix}
    F.~Landry, R.~Brock, P.~M.~Nadolsky and C.~P.~Yuan,
    Phys.\ Rev.\  D {\bf 67} (2003) 073016.

  \bibitem{Konychev:2005iy}
    A.~V.~Konychev and P.~M.~Nadolsky,
    Phys.\ Lett.\  B {\bf 633} (2006) 710.

                                              
  \bibitem{Sjostrand:2006za}
    T.~Sj\"ostrand, S.~Mrenna and P.~Skands,
    JHEP {\bf 0605} (2006) 026.


  \end{thebibliography}
\end{document}